\documentclass[aps,pre,twocolumn,groupedaddress,showpacs]{revtex4}
\pdfoutput=1

\usepackage{graphicx}
\usepackage{amssymb}
\usepackage{amsmath}
\usepackage{mathrsfs}
\usepackage{epstopdf}

\begin{document}

\title{Drying and cracking mechanisms in a starch slurry}

\author{Lucas Goehring}
\email[]{lg352@cam.ac.uk}
\affiliation{BP Institute for Multiphase Flow, Madingley Rise, Madingley Road,
Cambridge, UK, CB3 0EZ}

\date{\today}

\begin{abstract}
Starch-water slurries are commonly used to study fracture dynamics.  Drying starch-cakes benefit from being simple, economical, and reproducible systems, and have been used to model desiccation fracture in soils, thin film fracture in paint, and columnar joints in lava.  In this paper, the physical properties of starch-water mixtures are studied, and used to interpret and develop a multiphase transport model of drying.  Starch-cakes are observed to have a nonlinear elastic modulus, and a desiccation strain that is comparable to that generated by their maximum achievable capillary pressure.  It is shown that a large material porosity is divided between pore spaces between starch grains, and pores within starch grains.  This division of pore space leads to two distinct drying regimes, controlled by liquid and vapor transport of water, respectively.  The relatively unique ability for drying starch to generate columnar fracture patterns is shown to be linked to the unusually strong separation of these two transport mechanisms.   

\end{abstract} 


\maketitle

\section{Introduction}

\begin{figure}[h!]
\includegraphics[width=3.375in]{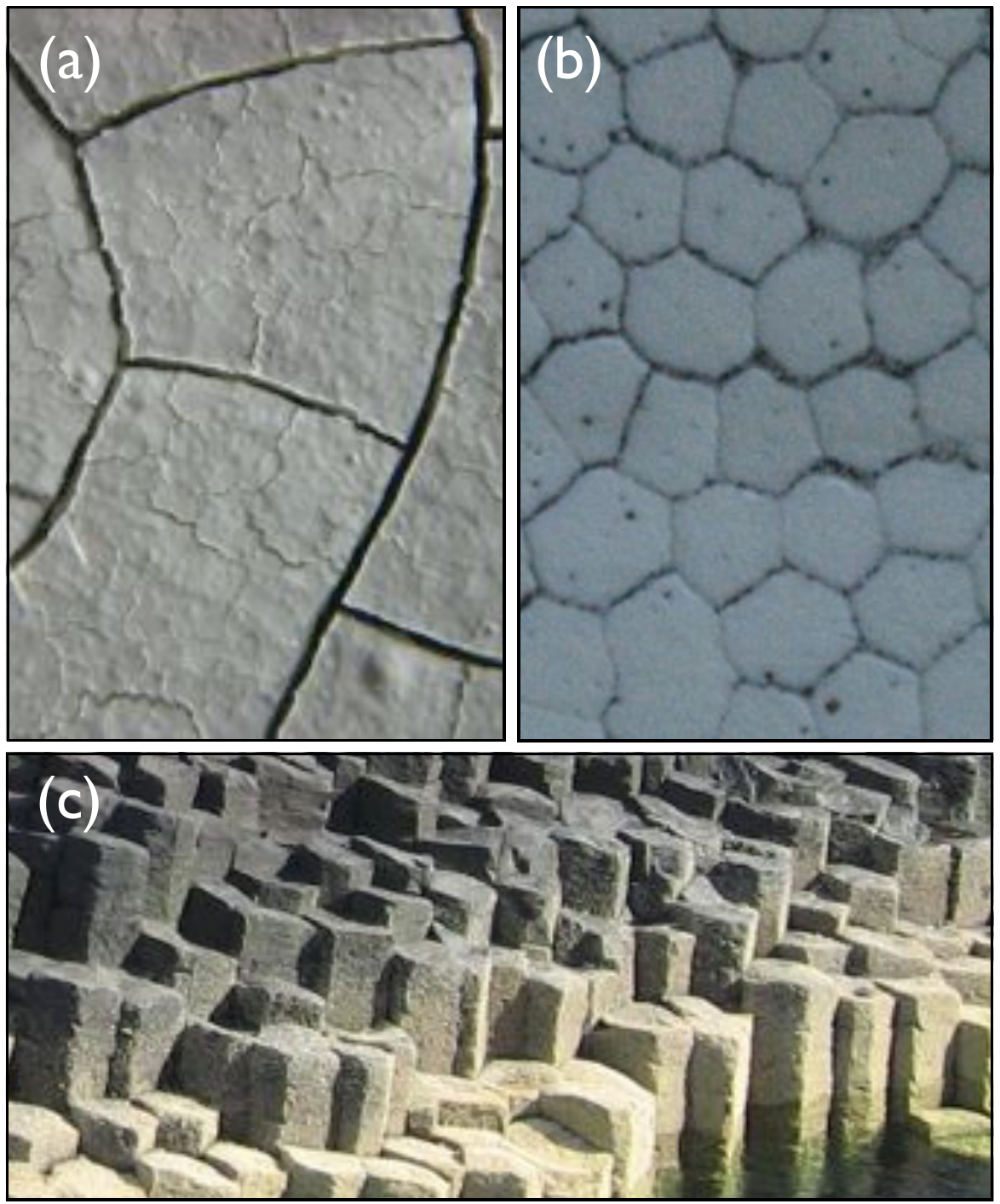}
\caption{\label{cracks} [Color online] Two types of cracks form in a drying starch-cake: (a) thin-film, or `mud-crack' patterns, and (b) columnar joints.  The thin-film crack spacing scales with film thickness \cite{Bohn2005,Bohn2005b}, and is typically a few cm.  The columnar joint crack spacing scales with the evaporation rate \cite{Goehring2006,Goehring2008}, and is typically around a mm. (c) Columnar joints are common in cooled lava, where they appear to be governed by similar scaling and organizational behavior to starch columns.}
\end{figure}

Mixtures of starch grains and water have attracted attention recently as model materials for studying drying and fracture patterns, and in particular the formation of columnar joints \cite{Muller1998,Muller1998b,Muller2000,Muller2001,Komatsu2001,Komatsu2003,Toramaru2004,Goehring2005,Mizuguchi2005,Bohn2005,Bohn2005b,Goehring2006,Nishimoto2007,Goehring2008,Goehring2009}.   When a starch-water slurry dries, it responds to desiccation in two distinct ways.  Initially, the evaporating slurry solidifies into a homogeneous starch-cake, which can be up to several cm thick, that dries uniformly throughout its thickness.  As shown in Fig. \ref{cracks}(a), stresses that occur in this phase drive the formation of a thin-film fracture pattern, similar to the craqulere patterns seen in pottery glazes or dried mud puddles \cite{Muller2000,Bohn2005,Bohn2005b}.  Further evaporation leads to a form of directional drying, as a desiccation front is initiated at the drying surface, and intrudes into the starch-cake \cite{Muller1998}.  This type of drying is associated with the development of columnar joints \cite{Goehring2009}, such as those shown in Fig. \ref{cracks}(b).  The resulting polygonal pattern is statistically identical to that found in columnar basalts, such as the Giant's Causeway \cite{Goehring2005}, or Fingal's Cave, shown in Fig. \ref{cracks}(c). Interestingly, the column size is also chosen by the same dimensionless scaling law in both cases \cite{Goehring2009}.  It remains mysterious, however, why drying starch slurries should form columnar joints at all.   Most other slurries or dispersions have only been seen to give rise to the more common craqulere patterns, when dried.  

Here, it is shown that the effective separation of the two very different drying mechanisms of liquid capillary flow and diffusive vapor transport is responsible for setting up the conditions of directional solidification that are necessary to produce columnar joints in a drying starch-cake. In the case of starch, this effective separation may arise from an unusually porous structure of the starch grains themselves.  

The stress state within a drying poroelastic solid is analogous to that of a cooling viscoelastic solid \cite{Biot1941,Norris1992}.  This correspondence allows experimental results obtained from drying starch to be compared, quantitatively, with the thermal problem of columnar joint formation in lava \cite{Goehring2008,Goehring2009}.  
Motivated by this well-known geophysical application, the dynamics of the columnar crack pattern have been studied experimentally in considerable detail \cite{Muller1998,Muller1998b,Muller2001,Toramaru2004,Goehring2005,Mizuguchi2005,Goehring2006,Goehring2008,Goehring2009}. Starch suspensions can also act as a conceptual bridge between colloids and soils, as the size of starch grains is intermediate between those of most soils, and those of common colloids.   There has been interest in studying water transport in starch as a soil analog \cite{Komatsu2003,Komatsu2001}, and to assist in explaining desiccation cracks in fine arid soils \cite{Ewing2006}.  The system has also attracted attention as a model test of simulations of crack formation \cite{Jagla2002,Jagla2002b,Jagla2004,Nishimoto2007,Tang2007}, as the ordered columnar joints are expected to arise spontaneously from the underlying dynamics.  In addition to quantiatively addressing the physical origin of columnar joints in drying starch, it is hoped that the full description of drying and cracking of starch presented in this paper will encourage such physically based modeling, and foster the development of a common understanding of the formation of of thin-film fracture, desiccation cracks in soils, and columnar joints.

The drying, and eventual fracture, of a slurry is a complex problem.  A detailed understanding of the physical properties of the network of grains and pores is necessarily involved in any effort to model desiccation, and columnar jointing.  Investigations were first made into several physical properties of desiccating starch slurries, in order to characterize how water may be transported either as a liquid or vapor, within an unsaturated starch-cake.  The results presented here are then used to develop and test a dynamical model of water transport, and of the microscopic origins of stresses, within a desiccating starch-cake, initially composed of equal masses of cornstarch and water. Finally, this model is used to suggest why starches, unlike most other porous media, form columnar fracture patterns when they dry.

\section{Physical properties of starch-water mixtures}

For all experiments presented here, cornstarch (sometimes known as cornflour) slurries were prepared by mixing equal masses of dry cornstarch (Canada Brand) and water in a glass container.  Starch-cakes, up to 4 cm thick, were prepared by pouring the initial slurry into 60 mm radius flat-bottomed glass dishes, which were then placed under a pair of 250 W heat lamps, to dry.  The materials and methods are identical to those reported in a number of studies of columnar jointing in starch \cite{Goehring2005,Goehring2006,Goehring2009}.

\subsection{Particle size}

The properties of granular porous media are strongly affected by the size and geometry of their consitiuent particles.  Using a laser-diffraction particle sizer (Malvern Mastersizer S), the average particle radius of starch grains in water was found to be $R_g$ = 8.2$\pm$0.1 $\mu$m, with the peak characterized by a full-width-half-maximum of 3.2$\pm$0.1 $\mu$m.  The starch grains do not swell when wetted -- 8 tests were performed, from immediately after mixing dry starch with water, to 15 minutes later, with no variation in size. 3 additional tests of grain size were done in air.  Some clumping of grains in air was noticed, but the position of the main peak of the particle size distribution was not displaced from that measured in water.  Particle size results were confirmed by the direct observation of a dilute sediment of starch in water, through a digital microscope.

\subsection{Particle surface character}

\begin{figure}
\includegraphics[width=3.375in]{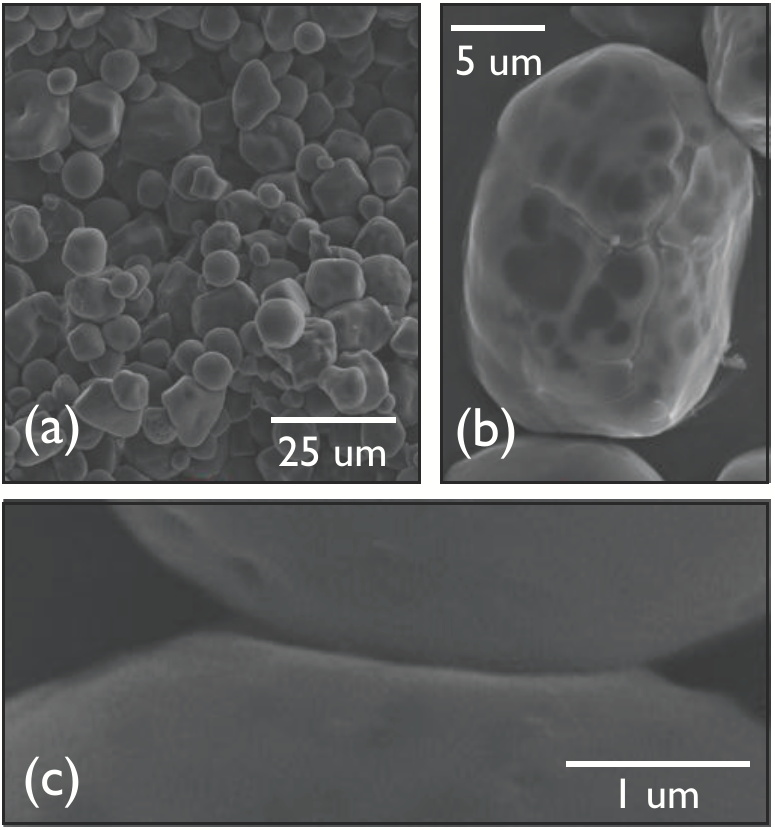}
\caption{\label{SEM} Scanning electron microscope images of starch. (s) Starch observed on the surface of a columnar joint are randomly, and closely packed.  (b) The starch grains are roughly spherical, and display both a platy, broken surface, and a highly porous sub-structure.  (c) Individual grain-grain contacts show deformation of the grains.}
\end{figure}

Electron microscopy was performed on a starch column selected from a fully dried columnar sample, and on dry starch grains dusted onto a sample holder.  The samples were sputtered with a thin layer of carbon, and observed in a JEOL JSM-840 scanning electron microscope.  When dried into a column, one surface of which is shown in Fig. \ref{SEM}(a), starch grains appear to be randomly close packed.  On close inspection, as shown in Fig. \ref{SEM}(b), the grain surface is somewhat platy and broken up, and the grains are seen to contain many pits.  When wetted these pits can potentially fill with water, which will need to be considered when estimating the porosity of any grain packing.  The large surface area of the pits, and their small diameter, may also account for some of the strong hydrophilic qualities of corn starch.   

If a starch suspension is allowed to dry and form columnar joints, the starch grains are damaged on the microscopic level.  If a sample is re-wetted, mixed, and re-dried, the resulting columns are usually larger and more irregular than those resulting from the first drying.  This damage is noticeable at grain-grain contacts in desiccated samples.  As shown in Fig. 2(c), these contacts involved the compression of one grain into the other, or the flattening out of sections of both grains.  It is likely that these `kissing' contacts were created by the action of water-bridges joining the neighboring grains.  Such bridges create a strong capillary force, pulling adjacent grains together.  Any strength of the final, desiccated columnar state is presumably due to the surface (van der Waals) forces acting across these contacts, as it is in powder compacts \cite{Kendall1988}.

\subsection{Rheology}

\begin{figure}
\includegraphics[width=3.375in]{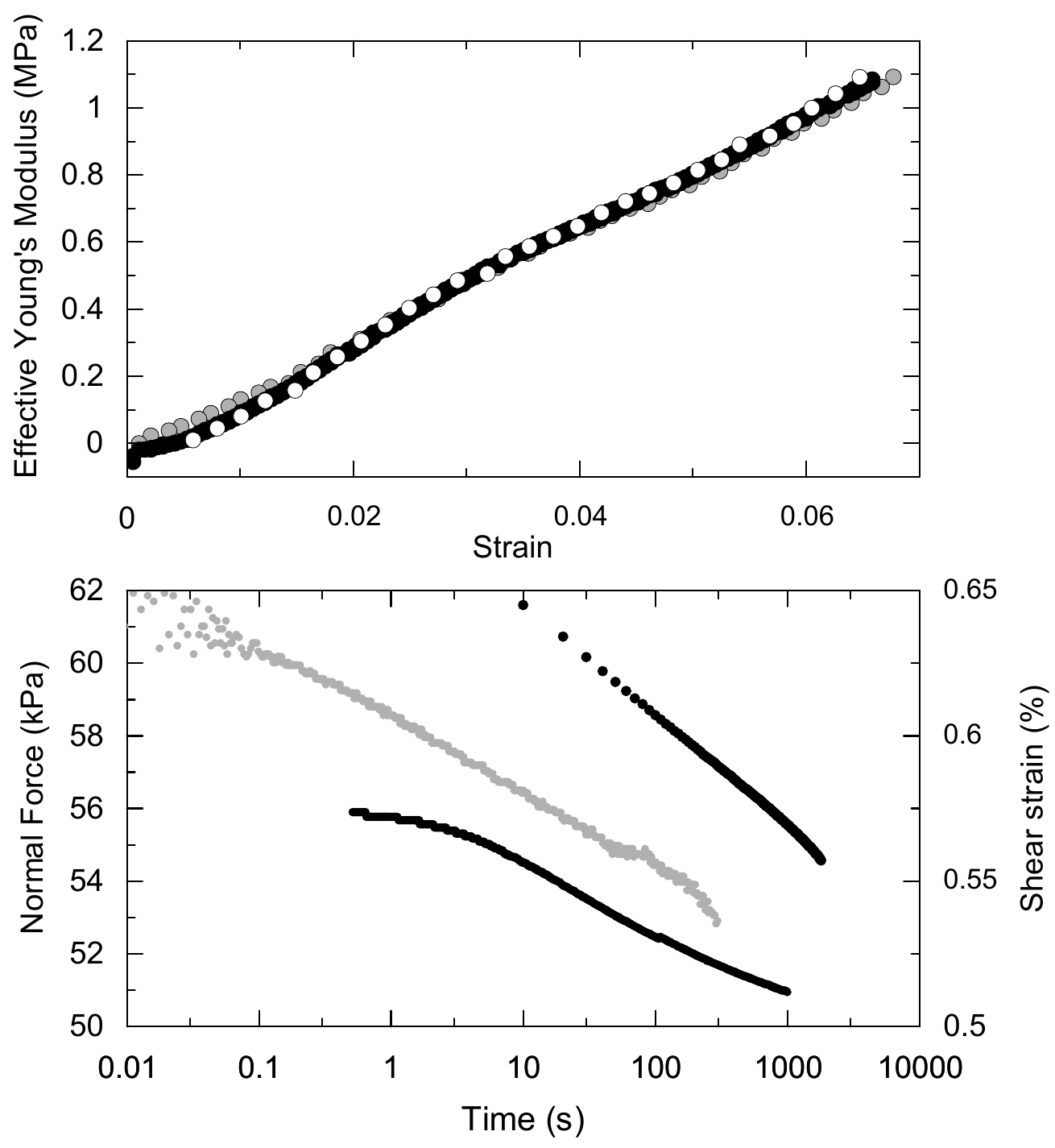}
\caption{\label{rheology} Drying starch exhibits highly non-lilnear rheology. (a) The effective Young's modulus in a 2 mm thick starch-cake of $C$ = 0.3 g/cm$^3$ is linearly dependent on the vertical strain (tests at strain rates of 0.1 $\mu$m/s (black), 0.5 $\mu$m/s (grey), and 2 $\mu$m/s (white) are shown). (b) In a starch-cake the logarithmic relaxation of compressional stress (black circles, left axis, two independent trials shown) and shear-strain (grey circles, right axis) continues for a long time after a stress or shear is applied.}
\end{figure}

Granular solids are known to have highly non-linear rheologies.  Since stress is only transmitted {\it via} particle-particle contacts, which behave as Hertzian springs \cite{Hertz1881}, a non-linear elastic model is appropriate.  The simplest such model to approximate a granular solid is that of 3rd order elasticity. This model was originally developed by Landau to describe anharmonic elastic effects in crystals  \cite{Landau1970}, but has been successfully applied to granular media in a wide variety of engineering and geophysical contexts \cite{Ostrovsky2001}.  In its general form this model adds the next-lowest (cubic) order terms to the elastic energy density.  In the absence of shear, the stress tensor
\begin{equation}
\sigma_{ii} = E \epsilon_{ii} + {\rm \tilde{C}} \epsilon_{ii}^2,
\label{elastic2}
\end{equation}
where $E$ is the Young's modulus, and ${\rm \tilde{C}}$ is one of three general Landau moduli, which describes the rate at which the stiffness of a body increases with increasing strain \cite{Landau1970}. The effective Young's modulus $E^\prime = \sigma_{ii}/\epsilon_{ii}$ was measured in a 2 mm thick, partially dried starch-cake by compressing and releasing the sample under a 40 mm diameter, parallel plate rheometer headpiece (TA Instruments RA1000), while observing the sample thickness and normal force response.   Experiments were performed over a range of compression and relaxation speeds, from 0.1-4 $\mu$m/s, and there was no observable effect of the strain rate on the normal force response.  As shown in Fig. \ref{rheology}(a), $E$, the elastic modulus at zero strain, was found to be consistent with zero, and the Landau modulus ${\rm \tilde{C}} $=16$\pm$1 MPa.  Tests at moisture concentrations of 0.3 and 0.4 g/cm$^3$ showed no difference in results.


An unusual feature of some granular solids is their lack of a well-defined stress relaxation time.  They can display a slow dynamical relaxation of an applied stress, that is logarithmic in time \cite{Ostrovsky2001}. This behavior belongs to a class of relaxation mechanisms shared by a number of superficially unconnected systems in energetically jammed states, such as of the magnetization of spin glasses, or the dc susceptibility of granular magnetic media \cite{TenCate2000}.  The causes of these slow dynamics are not particularly well understood, although Pauchard {\it et al.} have recently shown how the slow viscous deformation of nano-particles can relax stress \cite{Pauchard2009}.    The stress-relaxation of starch-cakes was tested by compressing a cake with a fixed strain, and monitoring the responding normal force over time.  Alternatively, a fixed shear stress was applied for 5 minutes, and the relaxation of the shear strain was monitored, after the stress was released.  As shown in Fig. \ref{rheology}(b), all starch-cakes showed a stress or residual shear that decayed logarithmically in time.  These results imply that stress relaxation will only play a small role of the rheological response of a starch-cake to stress; over the time-scale of a desiccation experiment, lasting of order 10$^4$-10$^5$ s, these relaxation mechanisms could only reduce an imposed stress by approximately 15-20\%.  Since the starch-cake remains irreversibly deformed after drying, however, some form of permanent particle deformation must occur.  This was observed to be the case at the kissing contacts shown in Fig. \ref{SEM}(c).

\subsection{Density}

\begin{figure}
\includegraphics[width=3.375in]{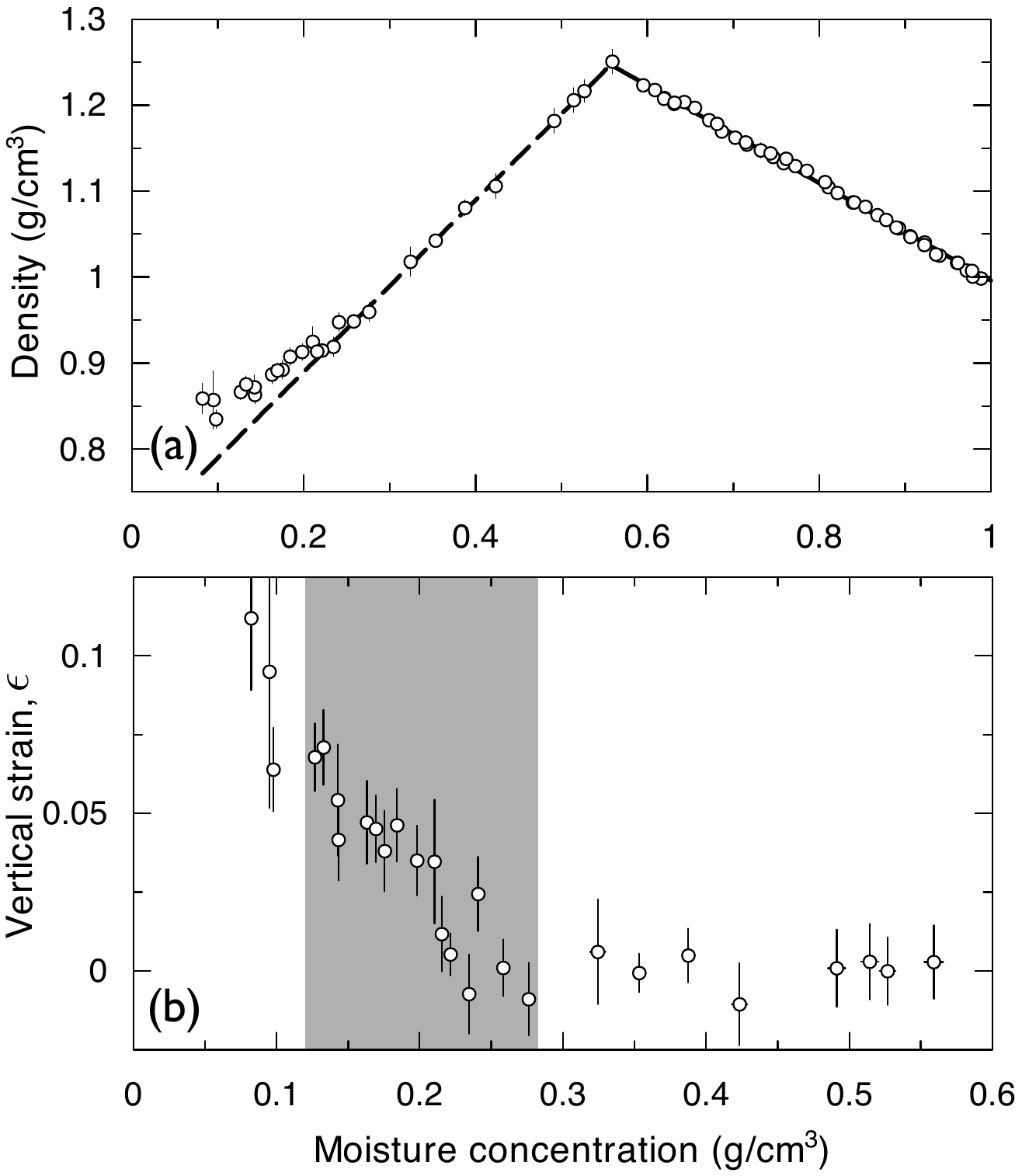}
\caption{\label{density} Density changes and strain develops during the drying of a starch slurry. (a) the measured density dependence (open circles) of starch-water mixtures on $C$ is well modeled by a non-interacting mixture of water, air, and starch (dashed curve).  (b) below $C$ = 0.3 g/cm$^3$ the density deviates from this curve due to desiccation-induced strain.  The grayed region indicates samples in which columnar joints were observed, but where joints did not reach the base of the starch-cake.}
\end{figure}  

The density of drying starch slurries was studied, in order to measure the porosity of a drying starch-cake, and to observe the onset and development of strain in the drying mixture.  Liquid starch-water mixtures were added to a 100 ml volumetric flask and weighed to determine their density.  An oscillating U-tube densitometer (Anton Paar DMA 500) was also used to measure fluid density, although for dilute samples the settling of starch out of suspension affected these measurements.  The densities of solid mixtures were calculated from measurements of the thickness, diameter, and mass of partially dried starch-cakes. 

If no tensile strain occurs during drying, then the dependence of the density $\rho$ of a starch-water mixture, on its water concentration $C$, is well fit by a simple model of non-interacting phases.  This mixture density, $\rho$, across a transition from a  two-phase (starch-water) mixture to a three-phase (starch-water-air) mixture, is described by
\begin{equation} \label{densitymodel}
\rho = \bigg\{ {(1-\rho_s/\rho_w)\rho_s+C, \atop (1-\phi) \rho_s + C,}  \mskip30mu {C \geq \phi \rho_w, \atop C \leq \phi \rho_w,}
\end{equation}
where $\rho_s$ is the bulk density of dry starch, $\rho_w$ is the density of water, and $\phi$ is the porosity of the packed starch-cake.  A fit of Eqn. \ref{densitymodel} to the density of starch mixtures, as shown in Fig. \ref{density}(a), yields $\rho_s$ = 1.57$\pm$0.01 g/cm$^3$, and $\phi$  = 0.56$\pm$0.01, and is in reasonable agreement with a previous measurement of the porosity of starch of $\phi$ = 0.51 \cite{Komatsu2003}.

Below a moisture concentration of approximately 0.3 g/cm$^3$, an unstrained three-phase density model is no longer valid, as the starch-cake responds to further drying by shrinking.  The average vertical strain in these samples was estimated by comparing the actual thickness of the starch-cakes with the expected unstrained thickness.  As shown in Fig. \ref{density}(b), the onset of strain is well correlated with the appearance of columnar jointing in the starch-cake.  The total strain measured is quite large, approximately 0.1.  Although some deformation must occur before the starch-cake dries to $C = 0.3$ g/cm$^3$, as primary cracks open during this phase of drying, the methods presented here could not detect the small strains involved.

\subsection{Permeability}

The transport of liquid water in a drying starch-cake occurs by flow through a porous medium.  In these conditions, Darcy's law
\begin{equation} \label{Darcy}
{\bf q}_l = -\frac{\kappa \rho}{\mu}\nabla P
\end{equation}
describes the mass flow rate per unit area, {\bf q}$_l$, of a fluid of viscosity $\mu$ and density $\rho$ that will pass through a porous body of intrinsic permeability $\kappa$, under a pressure gradient $\nabla P$.  The permeability was measured in order to characterize the efficacy of liquid transport during drying.  It is also used as a probe of the inter-particle porosity, $\phi_p$, which may differ from the total porosity $\phi$ = 0.56, due to the porous structure of the starch grains.  

A value of $\kappa =$(4.5$\pm$1.5)$\times$10$^{-14}$ m$^2$ was measured in saturated starch samples, at 24$^\circ$C.  A filter was glued to the bottom end of an open glass tube, and starch slurry was poured in from the top and allowed to settle.  Water was then added above the starch plug, and the flow rate through the starch was measured.  There was no dependence of the permeability on sample thickness or hydraulic head, and measurements were stable over long periods of time; one test was run for 80 hours, with no significant drift in the measured $\kappa$.

The Carmen-Kozeny equation
\begin{equation}
\label{permprop}
\kappa = \frac{R_g^2 \phi_p^3}{45(1-\phi_p)^2}
\end{equation}
is a commonly used semi-empirical model of the permeability of a porous medium \cite{Dufresne2003,Holland1995}.  Using a particle radius of 8.2 $\mu$m, and the measured permeability, Eqn. \ref{permprop} predicts that $\phi_p$ = 0.26$\pm$0.03, which suggests that the inter-particle porosity is much lower than the total porosity.  


\subsection{Water potential}

\begin{figure}
\includegraphics[width=3.375in]{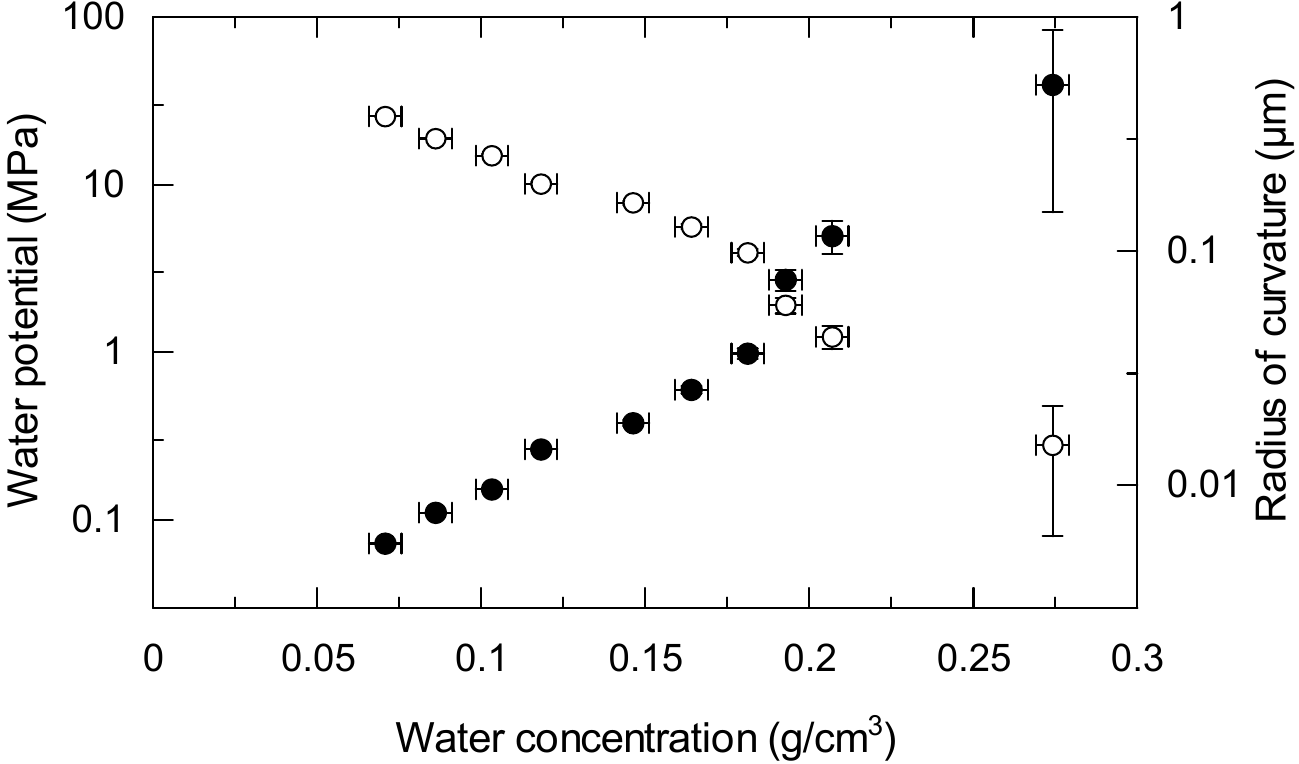}
\caption{\label{waterpot} The water potential (open circles) in partially dried starch-cakes increases as the cake dries.  The radius of curvature (filled circles) of water pockets in equilibrium with this potential drops dramatically as the cake dries.  Values above $C=$0.25 g/cm$^3$ are not within detectable limits. }
\end{figure}

In addition to fluid flow, water can be transported within a drying starch-cake by vapor diffusion.  These two mechanisms are in fact linked, as water can change phase. The equilibrium vapor pressure of water, $P_v$, over a liquid surface varies according to the Kelvin equation
\begin{equation}
\label{Kelvin}
P_v = P_{sat} e^{P \nu_m/RT},
\end{equation}
where $R$ is the gas constant, $\nu_m$ is the molar volume of water,  $P_{sat}$ is the equilibrium vapor pressure of water at temperature $T$ over a flat liquid surface , and the pressure drop
\begin{equation}
\label{tension}
P = -\frac{2\Gamma}{R_w}
\end{equation}
is caused by a surface tension $\Gamma$ acting across an air-liquid interface with a radius of curvature $R_w$ \cite{Morrison2002}.  For the concave menisci in a porous medium, $P_v$ is slightly reduced from $P_{sat}$.

The water retention curve (also known as the soil water characteristic curve) describes how $P$ changes with $C$, and can also be used to estimate the pore-size distribution of a material.  Shown in Fig. \ref{waterpot}, points along this curve were measured using a Decagon WPT4 water potentiometer.  1-2 mm thick starch-cakes were dried to a desired $C$, placed in the test cell, and allowed to equilibrate. $P$ was then measured by observing the relative humidity in the test cell.  Significantly above $C$ = 0.25 g/cm$^3$ the water potential was consistently below a detection limit of $\sim$0.5 MPa.  This is expected when $R_w$ is representative of the interparticle spaces, which are approximately an order of magnitude smaller than $R_g$ \cite{Lee2004}.  Below $C$ = 0.25 g/cm$^3$ the water potential rose rapidly as the starch-cakes dried, indicating that any remaining water is trapped in a distribution of much smaller pores.  

\subsection{Fluid phase}


25 g of starch were mixed with 500 ml of water for 1 hour using a stir bar. The resulting suspension was left to settle overnight, and decanted.  Measurements of the density of the supernate were indistinguishable from measurements of the density of tap-water.  Similarly, measurements of viscosity agreed with that of tap-water, below the gelation temperature $T_g$ = 62$^\circ$C.  Above $T_g$ the supernate's viscosity increased to approximately twice that of water, indicating a trace of dissolved amylose in solution \cite{Singh2002}.  These results confirm that, as long as experiments are kept below $T_g$, the fluid phase of a starch-cake can be treated as pure water.

\section{Drying dynamics}

\begin{figure}
\includegraphics[width=3.375in]{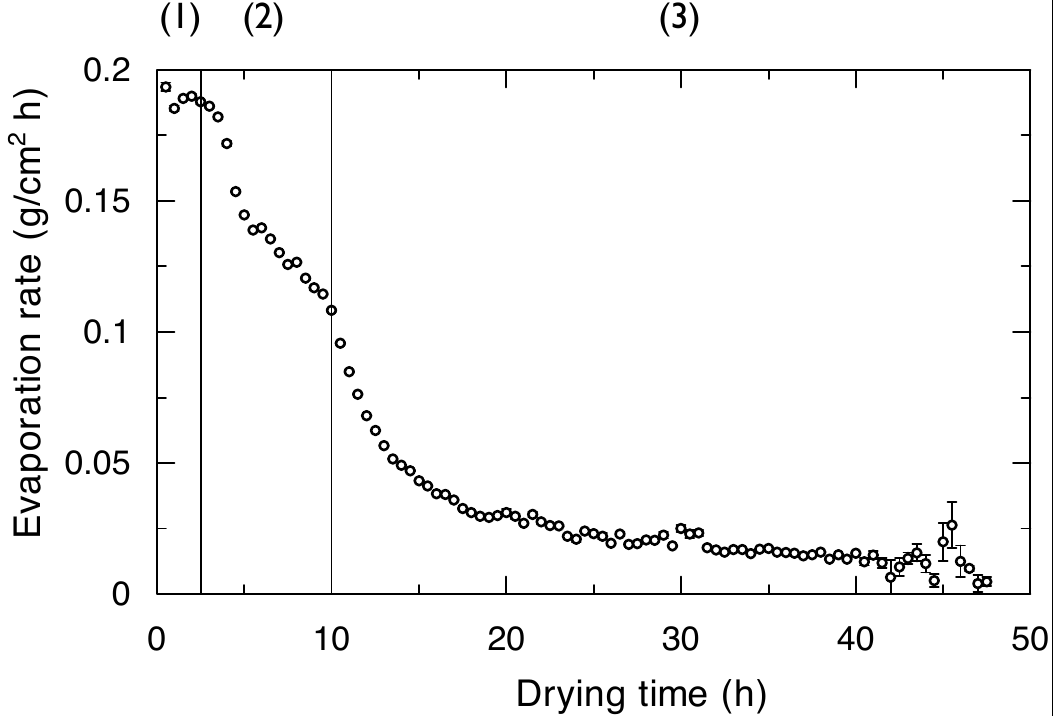}
\caption{\label{drying} The drying of a starch suspension, initially mixed with equal masses of water and starch, can be divided into three stages.  From the left to the right: (1) starch grains sediment in a water column, with surface evaporation; (2) a rigid mixture of starch, water, and air maintains a high surface evaporation, as capillary flow transports water to the drying surface; (3) once the capillary network becomes disconnected, the slower process of vapor transport completes drying.}
\end{figure}

The drying of a starch slurry occurs in three stages, as has been widely recognized \cite{Toramaru2004,Goehring2005,Mizuguchi2005,Bohn2005,Bohn2005b,Goehring2006,Nishimoto2007}. These are demonstrated in Fig. \ref{drying}.  Each phase is contolled by a different water transport process. 

Since most desiccation experiments begin with an excess of the fluid phase, so that the mixture can be manipulated as a liquid, the water content is initially higher than about 0.6 g/cm$^3$.  The first stage of drying occurs as the particles settle in a water column, while water evaporates from the surface of a pool of supernatant liquid.  The surface evaporation rate is constant, and has been used to characterize the relative drying rate of starch-cakes exposed to different drying conditions \cite{Toramaru2004,Goehring2006}.

\begin{figure}
\includegraphics[width=3.375in]{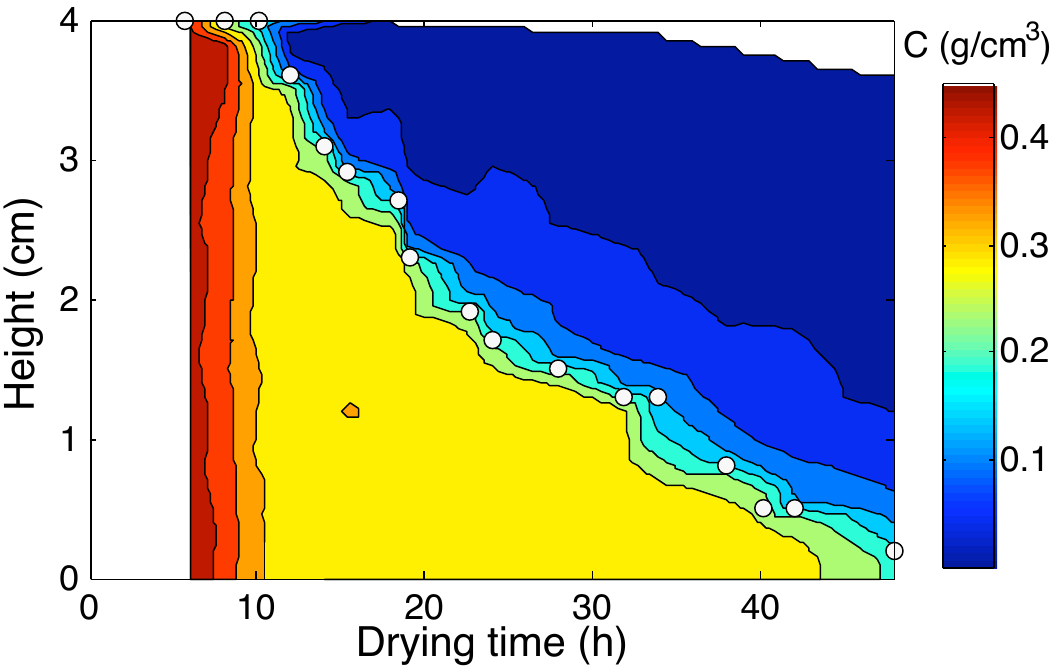}
\caption{\label{concentration} [Color online] Moisture concentration $C(z,t)$ measured in a series of identically prepared, partially dried starch-cakes.  Experiments were conducted to sample $C$ approximately every 3 h, and every 2 mm -- the displayed field is linearly interpolated between data, and contours are drawn every 0.05 g/cm$^3$.  The fracture front position (open circles) remains at $C_f$ = 0.20$\pm$0.01 g/cm$^3$, slightly trailing the beginning of the drying front.}
\end{figure}

In order to describe the dynamics of the later stages of drying, the water concentration $C(z,t)$ was sampled at different vertical heights $z$, measured from the base of the starch-cake, and times $t$ in a series of 17 identically prepared samples, each of which dried according to the evaporation rate shown in Fig. \ref{drying}.  Replicates were removed from the drying apparatus after different times $t$ and sliced into thin layers that were then weighed, fully dried, and reweighed in order to measure the water content.  The position of the columnar fracture front in each sample was also recorded.  The resulting moisture concentration field, shown in Fig. \ref{concentration}, sampled $C$ at roughly 3 hour intervals over 1-2 mm thick layers.  

When a starch-cake dries to $C$ = $\phi \rho_w$
, air begins to infiltrate into the voids between starch grains, initiating the so-called funicular regime of drying \cite{Lu2004}. This regime, also sometimes called the `first falling rate' period, is dominated by fluid flow through capillary bridges \cite{Lu2004}.    There is little to no measurable strain during this regime, although primary cracks do appear. By this time, the starch-cake is solid, has the consistency of moist chalk or clay, and is very weak and brittle.  The cake, however, still contains water equivalent to up to 56\% of its volume. A random close packing of spheres, which might be expected to roughly characterize the grain structure, has a porosity of only 36\%.  In order to realize such a large porosity, it appears likely that the water in a starch-cake is distributed between two distinct but connected reservoirs: the gaps left between the packed grains ($\phi_p$), and the smaller spaces within individual grains ($\phi_g$).
  
As evaporation progresses, larger pores will empty before smaller ones in equilibrium with them.   Thus, if there is a broad distribution of pore types, once most of the water has been removed from the pore space between grains, the remainder will be confined to small water bridges between particles, and any smaller pores within particles.  

As its water content falls past 0.30 g/cm$^3$, there is a sudden change in the drying behavior of the starch-cake. In similar cases, such a drop in the drying rate indicates a transition from a network of connected capillary bridges, to an unconnected network \cite{Lu2004,Nishimoto2007}.  At the onset of this  `pendular' regime, capillary flow abruptly ceases, and moisture can be transported only through the much slower means of vapor diffusion.  For a random close-packing of spheres, the funicular-pendular transition should occur at around $C$ = 0.05 g/cm$^2$ \cite{Flemmer1991}.  In order to explain the high value of $C$ observed at this transition, $\phi_g$ must contribute an additional porosity of approximately 0.25.  

During the pendular phase, a drying front forms, and passes through the sample, from top to bottom.  At this front, as shown in Fig. \ref{drying}, there is a clearly deliminated boundary between nearly uniformly moist unfractured starch, and starch which has given way to columnar joints.   In contrast to a typical diffusive front, the curvature $\partial^2C/\partial z^2$ is positive above the fracture front \cite{Goehring2006}.  Similar sharp drying fronts have been seen in the directional drying of 2D suspensions \cite{Dufresne2003}, and have been imaged in drying starch slurries using nuclear magnetic resonance techniques \cite{Mizuguchi2005}. Such a front is expected when there is a shift from flow-limited to diffusion-limited water transport \cite{Dufresne2003,Goehring2006,Nishimoto2007}.

\section{Water transport model}

In order to quantify the remarks made in the previous section, a model of water transport is briefly outlined here.  The basic model was recently proposed by Nishimoto {\it et al.} \cite{Nishimoto2007}, and further details are given there. The model presented here has been modified to account for the implications of pore space within starch grains.

Fluid flow in a porous structure is governed by Darcy's law, Eqn. \ref{Darcy}, with a permeability $\kappa(C)$ that depends on the degree of saturation.  Expressed in terms of a concentration gradient, this implies that the fluid flux
\begin{equation} \label{captrans}
{\bf q}_l = -\frac{\kappa(C) \rho_w}{\mu} \frac{\partial P}{\partial C} \nabla C.
\end{equation} 
Here, variations in the driving pressure $P$ arise from the dependance of the radius of curvature of the capillary bridges connecting particles, on $C$.

The transport of water vapor is diffusive.  Since the equilibrium vapor concentration is proportional to the vapor pressure $P_v$, the Kelvin equation, Eqn. \ref{Kelvin}, can be used to show that
\begin{equation}
\label{diffusion2}
{\bf q}_v = - \bigg( \frac{D_0 C_{sat} \nu_m}{\tau RT}  (\phi - C/\rho_w) e^{P \nu_m/RT}  \frac
{\partial P}{\partial C}\bigg) \nabla C,
\end{equation}
where $C_{sat}$ is the saturated concentration of water in air, $D_0$ is the free diffusion constant of water vapor in air, and $\tau$ is the tortuosity of the network through which the gas must diffuse \cite{Nishimoto2007}.  

\begin{figure}
\includegraphics[width=3.375in]{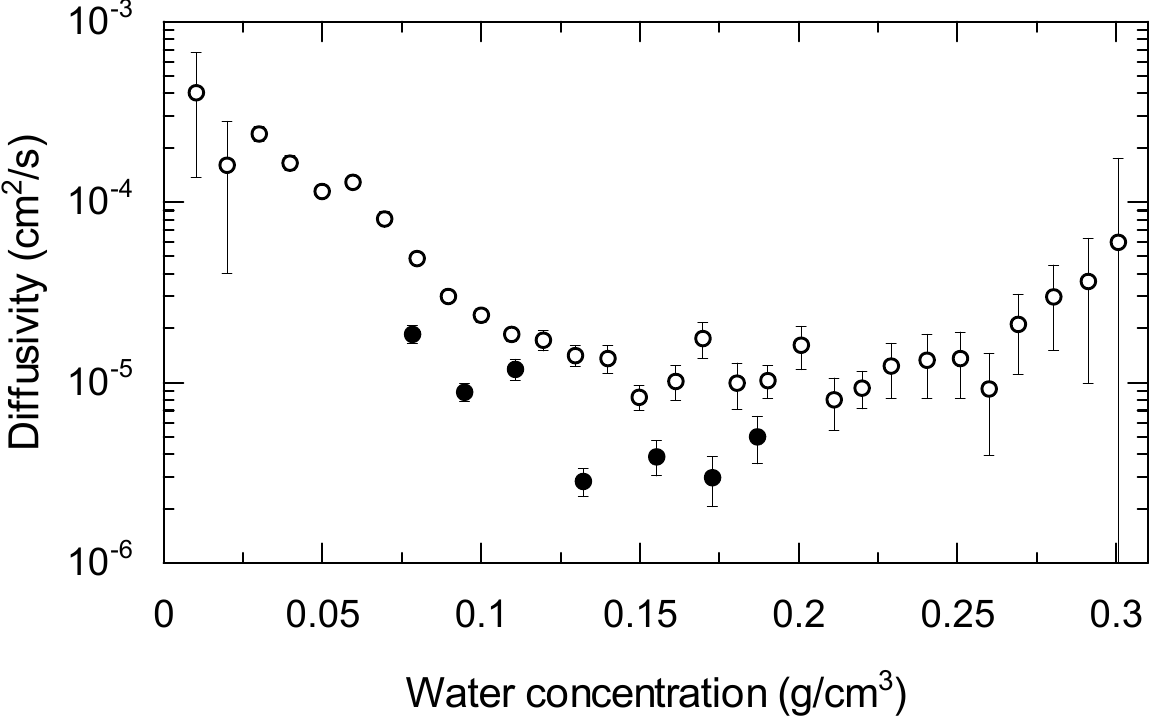}
\caption{\label{DFig}  The variation of the diffusivity of water within a drying starch cake can be measured by either considering the evolution of the concentration (Eqn. \ref{DofC}, open circles), or the water potential (Eqn. \ref{Diffusivity}, filled circles).  The water potential data assume that $D_0$ = 0.3 cm$^2$/s and $C_{sat}$ = 5.1$\times$10$^{-5}$ g/cm$^3$ are taken at a typical temperature $T$ = 40$^\circ$C of the starch-cake, that $\tau$ = $\sqrt{2}$, $\rho_w$ = 1 g/cm$^3$, $\phi$ = 0.56, and that $V_w$ = 18 cm$^3$/mol. Above $C = 0.3$ g/cm$^3$, $\partial C/ \partial z$ is small, and Eqn. \ref{DofC} cannot be accurately calculated. }
\end{figure}

By combining the liquid and vapor fluxes with a mass conservation equation, one obtains a nonlinear diffusion equation
\begin{equation}
\label{nonlin}
\frac{\partial C}{\partial t}= -\nabla \cdot ({\bf q}_l + {\bf q}_v) = \nabla \cdot (D(C) \nabla C),
\end{equation}
which has previously been used to describe mass transport in a drying slurry \cite{Mizuguchi2005,Goehring2006,Nishimoto2007,Goehring2009}.  In this case, the diffusivity
\begin{equation}
\label{Diffusivity}
D =  \bigg( \frac{D_0C_{sat} \nu_m}{\tau RT}  (\phi - \frac{C}{\rho_w}) e^{P \nu_m/RT}  + \frac{\kappa(C) \rho_w}{\mu} \bigg)  \frac{\partial P}{\partial C}
\end{equation}
Alternatively, by integrating Eqn. \ref{nonlin}, and by assuming that the concentration varies only over the vertical position $0 \le z \le h$, $D(C)$ can be calculated from the measurements of the concentration field presented in Fig. \ref{concentration} \cite{Pel2002,Goehring2009},
\begin{equation}
D(C(z,t)) = \bigg({\int_0^z \frac{\partial C}{\partial t} d z^\prime}\bigg) \bigg/ \bigg({\frac{\partial C}{\partial z}}\bigg).
\label{DofC}
\end {equation}

Evaluating Eqn. \ref{nonlin} requires considerable detailed knowledge of the system under study. In the funicular regime, there exist a number of semi-empirical models of capillary transport \cite{Pel2002,Lu2004,Nishimoto2007}, which typically assume either a power law \cite{Nishimoto2007} or exponential \cite{Pel2002} increase in the diffusivity with increasing water concentration.  In Fig. \ref{DFig} the measured concentration field has been used to calculate $D(C)$.  These data do not discriminate between the competing empirical models, but demonstrate how the diffusivity increases rapidly between $C = 0.25$ and 0.3 g/cm$^3$. This suggests that at this water concentration, the starch-cake is indeed reaching the pendular-funicular transition, with an estimated $\phi_g$ = 0.25.

In the pendular regime, when the fluid network is unconnected, it is assumed that all moisture transport occurs within the vapor phase, and that $\kappa$ = 0.  Under this assumption, as shown in Fig. \ref{DFig}, the water potential $P(C)$ can be used to estimate the diffusivity $D(C)$ directly, using Eqn. \ref{Diffusivity}. This interpretation of the water potential is compared to the solution of Eqn. \ref{DofC} in Fig. \ref{DFig}, and accurately captures both the beginning of an rapid rise in $D(C)$ at low moisture concentrations, and a plateau in $D(C)$ at intermediate concentrations.  The small discrepancy, approximately a factor of two, between $D(C)$ as calculated from Eqns. \ref{Diffusivity} and \ref{DofC} may be attributable to diffusive transport through the starch grains, to local transport of water within pores, or to temperature gradients within the starch-cake ($C_{sat}$ is strongly temperature-dependant), which are not included in the model.  The reasonable agreement between the two methods implies that Eqn. \ref{diffusion2} accurately describes a vapor-controlled transport of moisture in the drying starch-cake, below $C\simeq$ 0.20 g/cm$^3$.

The transition between liquid and vapor transport mechanisms is characterized by a minimum in $D(C)$, which creates a bottleneck in the water transport to the drying surface. The existence of some minimum in $D(C)$ has been described \cite{Goehring2006,Nishimoto2007,Goehring2009}, but a broad, deep minimum appears to be necessary for creating the sharp drying front associated with the columnar fracture front \cite{Goehring2009}.  The depth and position of this minimum depends on the water storage capacity of the starch grains.  In the absence of appreciable pore space within grains, the pendular regime would be confined to $C\leq 0.05$ g/cm$^3$ \cite{Flemmer1991}.  Under these conditions, Nishimoto {\it et al.} used a semi-empirical model, adapted from soil mechanics, to evaluate Eqn. \ref{DofC}.  They predicted a minimum in the diffusivity that was one-to-two orders of magnitude larger than observed \cite{Nishimoto2007}. In this case, the resulting moisture distribution, $C(z,t)$, would be considerably more diffuse.  

\section{Fracture mechanics}

There are two types of fracture that have been studied in drying starch-cakes.  While the cake is in the pendular regime there is effectively no variation of water concentration on position, and first-generation (also known as primary) cracks appear that penetrate the full depth of the drying cake.  Later, during the funicular regime, drying is essentially confined to a thin interface between the wet and dry regions of the starch-cake. Columnar joints form during this phase, with the crack tips following the drying front.

The plumose structure on the primary cracks show that they initiate at the upper drying surface, and quickly propagate vertically through the entire sample \cite{Muller2000}.  The resulting disordered fracture networks scale like those of ordinary thin films, and have been studied as examples thereof \cite{Bohn2005,Bohn2005b}.   In particular, the average spacing between first generation cracks increases roughly linearly with sample depth, and these cracks typically meet at 90$^\circ$ junctions \cite{Bohn2005,Bohn2005b}.

As it enters the funicular regime, a starch-cake begins to dry from the top down. Water is confined in unconnected reservoirs within particles, and in inter-particle capillary bridges.  During the final phase of drying, a total strain of up to 0.1 develops, which, according to Eqn. \ref{elastic2}, is equivalent to putting the sample under a uniform bi-axial stress of up to 160 kPa (the vertical stress can be relieved by lowering the drying surface). Internally, this strain is accommodated by the deformation of grains into each other, as shown in Fig. \ref{SEM}, through the process of liquid sintering \cite{Lampenscherf2000}.  This measured stress is in reasonable agreement with the expected maximum attainable capillary pressure, typically taken to be of order 12$\gamma$/$R_g$ = 110 kPa  \cite{Lee2004,Dufresne2003}.  

A fine-scale fracture pattern develops at the drying surface at the start of the funicular regime, and further drying propagates the crack tips into the sample.  This leads to a moving fracture front that tracks a clearly delineated boundary between relatively soft, moist starch, and harder dry starch, which has given way to columnar jointing.  These two layers come apart easily when a partially dried sample is removed from its dish, as all the fracture tips are confined to a thin interface.  

The directional drying of an unsaturated starch-cake is related to the directional drying of thin colloidal layers \cite{Allain1995,Dufresne2003,Gautier2007}. In these experiments, colloidal dispersions are typically confined within a small cell, and allowed to dry from one exposed face.  Stress builds up in the drying body, in response to a gradient in pore pressure.  Eventually, an array of cracks form at the exposed edge, and slowly advance into the solidifying network.  Like columnar joints, these cracks are evenly spaced out, and advance perpendicular to a drying front.  However, unlike joints in drying starch, it should be pointed out that cracks in colloidal dispersions occur while the drying body is still saturated with water \cite{Dufresne2003}.    

Throughout unsaturated drying, the surface tension of the water bridges between grains gives a partially dried starch-cake much of its cohesiveness, and acts to pull the starch grains together.  The bridge-grain structure has been modeled as a spring whose equilibrium position varies with $C$, and which breaks when stretched beyond finite displacement \cite{Komatsu1997,Jagla2002,Nishimoto2007}. The compaction of grains into each other leads to the development of cracks, as any set of particle-particle contacts will contain some bonds which are weaker than others.  By allowing such bonds to yield, the surrounding network of bridges relaxes.  These assumptions give rise to fracture patterns that agree qualitatively with experiments, although there has been some difficulty in studying model systems sufficiently large enough to be checked quantitatively against experimental observation \cite{Komatsu1997,Jagla2002,Nishimoto2007}. Future developments of this type of discrete-element model would ideally couple the water transport dynamics described above with the evolution equations for the springs.

Columnar joints have been reported in many thermoelastic systems, such as lava, smelter slag, and optical glass \cite{Degraff1987}.  However, starch remains the only known system in which columnar joints have been studied in detail.  Indeed, this unusual feature of dried starch-cakes has been independently rediscovered on several occasions \cite{Huxley1881,French1925,Muller1998}. It has recently been shown that the diameter of columnar joints scales with the thickness of the drying front \cite{Goehring2009}, which is in turn controlled by the cross-over in drying dynamics from flow-dominated transport, to vapor-dominated transport.  In drying starches, this front is typically a few mm wide \cite{Goehring2009}, and columnar joints are a few mm in diameter.   

It has been argued that columnar jointing occurs when the length scale set by the minimum diffusivity, and a characteristic drying time, is smaller than the thickness of the drying layer \cite{Goehring2009}. In the previous section, it was demonstrated that the sudden transition between wet and dry starch arises from a broad, deep minimum in the diffusivity of moisture throughout the starchcake, which has been interpreted to arise from a large internal porosity, within starch grains.  In the absence of these pores, the minimum diffusivity has been estimated to be significantly larger \cite{Nishimoto2007}, and the drying front would have a characteristic width of many cm.  This suggests why, in most natural or laboratory situations, given that most soils do not share the observed porous structure of starch grains, the evolution of desiccation fracture patterns typically halts at the mud-crack stage.   In other cases it may be possible to produce columnar joints, if the drying body is large enough.  The experiments of Gauthier {\it et al.} \cite{Gautier2007}, who observed the slow propagation of two perpendicular cracks, meeting at a vertex, in colloidal silica confined within a 1 mm diameter tube, suggest that 3-dimensional crack patterns such as columnar joints may be experimentally accessible in other drying systems of modest scale, for example.

\section{Conclusion}

Starch-water mixtures shrink and crack when dried.  The simplicity of this system, and its use as a model for studying columnar joints, have encouraged considerable interest in the drying and cracking behavior of starch-cakes \cite{Muller1998,Muller1998b,Muller2000,Muller2001,Komatsu2001,Komatsu2003,Toramaru2004,Goehring2005,Mizuguchi2005,Bohn2005,Bohn2005b,Goehring2006,Nishimoto2007,Goehring2008,Goehring2009}.  

The results presented here have shown, quantitatively, how directional drying occurs in starch-water mixtures. An effective separation of vapor and liquid water transport creates a bottleneck in the drying behavior, leading to this type of drying.  

In order to evaluate how water is transported through a starch-cake, several properties of starch were measured.  An unusually large cake porosity was observed, which was argued to be divided roughly equally between pore space within grains, and pore space between grains.  By adapting a model of moisture transport, it was shown how this division affects the transition between flow-dominated, and vapor-dominated water transport, causing the liquid transport to become ineffective in circumstances where considerable water remains in the body.  A broad, deep minimum in the effective diffusivity of moisture followed, and lead to a sharp, propagating drying front. The total strain associated with this front is comparable to that predicted by capillary effects between grains.  Such a sharp shrinkage front appears to be necessary to generate columnar joints \cite{Goehring2009}.   The relatively unique behavior of drying starch slurries to give rise to columnar joints can thus be seen to arise directly from the unusual porous structure of starch grains, and the mechanics of moisture transport. 

\section{Acknowledgments}
The author wishes to thank S. W. Morris and A. Nishimoto for fruitful discussions, and A. F. Routh for a critical reading of an early draft, and is grateful for access to apparatus across several departments at the University of Toronto, and the University of Washington.

\end{document}